\documentclass[journal]{vgtc}                % final (journal style)
\ifpdf%                                % if we use pdflatex
  \pdfoutput=1\relax                   % create PDFs from pdfLaTeX
  \usepackage{graphicx}                % allow us to embed graphics files
  \DeclareGraphicsExtensions{.pdf,.png,.jpg,.jpeg} % for pdflatex we expect .pdf, .png, or .jpg files
\else%                                 % else we use pure latex
  \usepackage{graphicx}                % allow us to embed graphics files
  \DeclareGraphicsExtensions{.eps}     % for pure latex we expect eps files
\fi%

\graphicspath{{figures/}{pictures/}{images/}{./}} % where to search for the images

\usepackage{microtype}                 % use micro-typography (slightly more compact, better to read)
\PassOptionsToPackage{warn}{textcomp}  % to address font issues with \textrightarrow
\usepackage{textcomp}                  % use better special symbols
\usepackage{mathptmx}                  % use matching math font
\usepackage{times}                     % we use Times as the main font
\usepackage{cite}                      % needed to automatically sort the references
\usepackage{tabu}                      % only used for the table example
\usepackage{booktabs}                  % only used for the table example
\usepackage{gensymb}
\usepackage{textcomp}
\usepackage{enumitem}
\usepackage{balance}
\usepackage{setspace}
\usepackage[usenames,dvipsnames]{xcolor}

\newcommand{\pie}{pie chart task}

\onlineid{0}

\vgtccategory{Research}
\vgtcpapertype{evaluation}

\title{Professional Differences: A Comparative Study of Visualization Task Performance and Spatial Ability Across Disciplines}

\author{Kyle Wm. Hall, Anthony Kouroupis, Anastasia Bezerianos, Danielle Albers Szafir, and Christopher Collins}
\authorfooter{
\item
 Kyle Wm. Hall was with Temple University during this work. \newline Present Affiliation: Layer 6,  E-mail: kyle@layer6.ai.
\item
 Anthony Kouroupis and Christopher Collins are with Ontario Tech University. E-mail: anthony.kouroupis@ontariotechu.net, christopher.collins@ontariotechu.ca.
\item
Anastasia Bezerianos is LRI-Université Paris-Saclay. E-mail: anab@lri.fr.
\item
 Danielle Szafir is with University of Colorado Boulder. E-mail:
danielle.szafir@colorado.edu.
}

\shortauthortitle{Hall \MakeLowercase{\textit{et al.}}: Spatial Perception Differences by Profession}

\abstract{Problem-driven visualization work is rooted in deeply understanding the data, actors, processes, and workflows of a target domain. However, an individual's personality traits and cognitive abilities may also influence visualization use. Diverse user needs and abilities raise natural questions for specificity in visualization design: \textit{Could individuals from different domains exhibit performance differences when using visualizations? Are any systematic variations related to their cognitive abilities?} This study bridges domain-specific perspectives on visualization design with those provided by cognition and perception. We measure variations in visualization task performance across chemistry, computer science, and education, and relate these differences to variations in spatial ability. 
We conducted an online study with over 60 domain experts consisting of tasks related to pie charts, isocontour plots, and 3D scatterplots, and grounded by a well-documented spatial ability test. Task performance (correctness) varied with profession across more complex visualizations (isocontour plots and scatterplots), but not pie charts, a comparatively common visualization. We found that correctness correlates with spatial ability, and the professions differ in terms of spatial ability. 
These results indicate that domains differ not only in the specifics of their data and tasks, but also in terms of how effectively their constituent members engage with visualizations and their cognitive traits. Analyzing participants' confidence and strategy comments suggests that focusing on performance neglects important nuances, such as differing approaches to engage with even common visualizations and potential skill transference. Our findings offer a fresh perspective on discipline-specific visualization with specific recommendations to help guide visualization design that celebrates the uniqueness of the disciplines and individuals we seek to serve.}


\teaser{
  \centering
  \includegraphics[width=\linewidth]{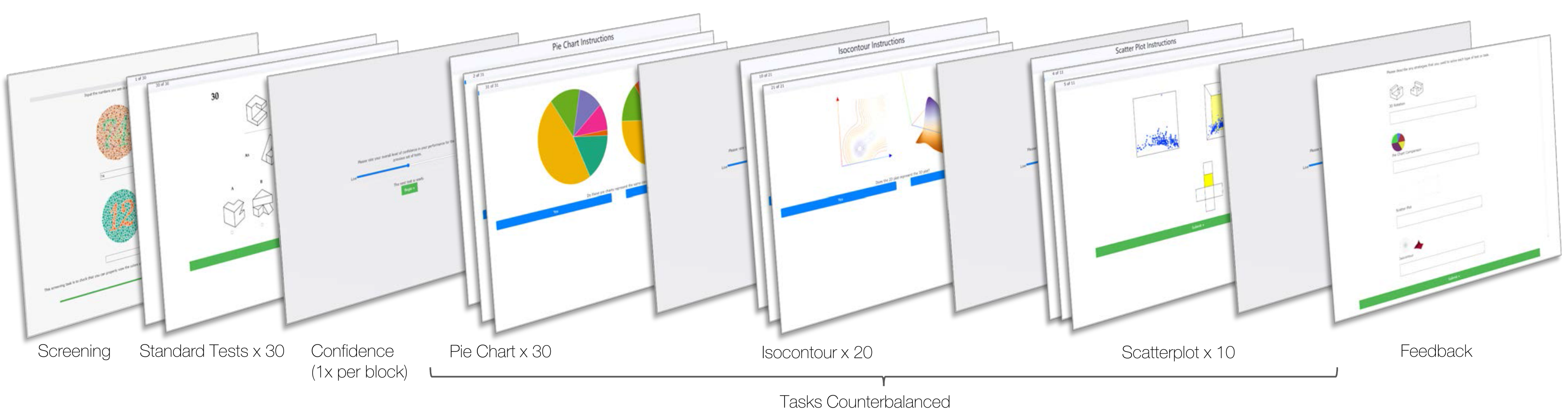}
  \caption{We conducted a spatial abilities study across three professional populations. After screening, demographics, and a standardized set of spatial reasoning tasks, participants carried out three task blocks in counterbalanced order, followed by feedback on their problem-solving strategies.}
  \label{fig:teaser}
}

\vgtcinsertpkg

\begin{document}


\firstsection{Introduction}

\maketitle

Empowering individuals to understand their data is the enduring promise of the multidisciplinary venture that is visualization. While the approaches, materials, and philosophical perspectives of visualization 
continue to evolve, the goal of visualization remains largely the same: 
\textit{how can we best serve individuals seeking to understand their data?} 
Visualization research has addressed this question following 
several lines on inquiry, including
\textbf{1)} how people ``see'' data and \textbf{2)} how disciplines differ in terms of their data and tasks. The results of this work 
are foundational
design guidelines grounded in
the perceptions and reasoning processes of the monolithic ``user'' and discipline-specific characterizations of target problem spaces supporting 
nuances in design. Many have leveraged this pragmatic dichotomy to great success. Once we understand the data and discipline, we essentially rely on universal guidelines grounded in the generic ``user'' to create the right design.  

While separating the discipline and data from the user simplifies design, prior studies have shown that reality is more complex: individuals have different cognitive abilities and those abilities inform visualization use \cite{VelezEtAlVis2005,wu2021understanding,PeckEtAlBELIV2012}. 
Educators have explored in detail the relationships between cognitive abilities, educational path, and profession \cite{WaiEtAl2009,SheaEtAl2001,WebbEtAl2007,YoonMetApp2016}. However, we do not know how these two elements---profession and cognitive traits---interact when using visualizations. 
Should designers characterize the cognitive abilities of a target discipline as part of the design process, or is one-size-user-fits-all cognitive guidance about effective visualization designs sufficient?

We explore the relationship between profession, cognitive traits, and visualization use through the lens of variation in \emph{spatial abilities}, defined as ``the ability to generate, retain and manipulate abstract visual images''~\cite[p. 188] {Lohman1979}. Spatial abilities influence how effectively people use visualizations, including their strategies for exploring data and abilities to draw relations across that data \cite{ChenCZerwinski1997,VicenteEtAl1987,OttleyEtAlVIS2015,CohenHegarty2007,VelezEtAlVis2005}. However, prior training and experience can likewise influence an individual's abilities to use visualizations, even for people with low spatial abilities \cite{froese2013evaluation}. 
We measured the associations between visualization performance, spatial ability, and discipline in an online study 
across computer scientists, chemists, and educators. Our results reveal that visualization task performance exhibits a substantive 
interplay between discipline and spatial ability, revealing
limitations for the universal user approach employed by current visualization design processes. We provide a new lens for understanding visualization based on \textit{professional cognitive differences}~\textemdash~systematic variations in disciplinary visualization usage concomitant with cognitive variations. 

In this paper, we contribute: 
\textbf{1)} detailed statistical analyses demonstrating that visualization task performance (both correctness and completion times) varies with both discipline and spatial ability, 
\textbf{2)} a demonstration that enhanced correctness can coincide with other behavior (e.g., the perceived importance of graphical representations in the workplace), and
\textbf{3)} a qualitative analysis of self-reported strategies, highlighting the importance of discipline-specific skills and concepts transferring to general tasks. 
Our results indicate that the expediency and efficiencies of prevailing disciplinary engagement and design models within visualization may have a hidden cost: designing for the universal user may obfuscate the full richness of whole persons whose professional choices and cognitive abilities are not easily separated.

\section{Related Work}
\label{sec:background}

Our study bridges perspectives on visualization use across both problem-centered (i.e., domain expertise) and user-centered (i.e., user cognition) perspectives. We draw on prior knowledge and methods from across these fields to investigate the relationship between spatial abilities, professional discipline, and visualization comprehension. 

\subsection{Cognitive Visualization Studies}
Visualization has a rich literature investigating psychological and cognitive factors associated with visualization use. This work typically focuses on understanding what elements of design make visualization generally useful for the typical user. However, this broad lens fails to consider nuanced differences emerging when any given person uses a visualization, such as ability levels \cite{froese2013evaluation,VanderPlasHofmann2016}, prior knowledge \cite{xiong2019curse}, data literacy \cite{mansoor2018data}, and personality traits \cite{ottley2015personality}.  
Moreover, such work is also counterbalanced by a growing body of work relating variations in visualization performance and usage to cognitive differences between individuals, otherwise known as \textit{individual differences}~\cite{LiuEtAlCGF2020}. In their framework characterizing individual differences for visualization, Peck et al.~\cite{PeckEtAlBELIV2012} proposed that cognitive differences between individuals could be classified according to three orthogonal dimensions:
\vspace{-0.5em}
\begin{itemize}[topsep=4pt,parsep=1pt,itemsep=3pt]
\item \textbf{Cognitive State}, which captures the instantaneous mental state of a person (e.g., in terms of cognitive load and emotions); 
\item \textbf{Cognitive Traits}, which are relatively stable over a person's adult life and include both personality traits (e.g., extroversion) and cognitive abilities (e.g., spatial ability); and  
\item \textbf{Experience/Bias}, which embodies the influence of a person's previous interactions and experiences. 
\end{itemize}
\vspace{-0.5em}
By treating cognitive traits and experience/bias as orthogonal, this framework implicitly asserts that an individual's cognitive traits do not influence the experiences that individual pursues (e.g., education and profession) and vice versa. This decoupling is also present in visualization studies on individual differences where participant's professional information is typically either: \textbf{1)} not reported~\cite{VicenteWilliges1988,VelezEtAlVis2005, OttleyEtAlVIS2015}, or \textbf{2)} partially reported, but not analyzed as a potential factor influencing performance~\cite{ChenCZerwinski1997,Chen2000,CohenHegarty2007}. Studies~\cite{ChenCZerwinski1997,Chen2000,CohenHegarty2007} that do report professional information generally employ a professionally homogeneous participant pool, failing to consider the nature of professional cognitive differences and their relation to visualization. We next review work related to spatial ability, the cognitive trait we focus on in this study. 

While domain-focused participatory methodologies imply that an individual's professional discipline can impact the right design for a given domain~\cite{SedlmairEtAlTVCG2012,HallEtAlTVCG2020}, previous visualization work~\cite{ChenCZerwinski1997,VicenteEtAl1987,luo2019user} has demonstrated that 
spatial ability has a complex influence on visualization use and preference.
For example, Chen \& Czerwinski~\cite{ChenCZerwinski1997} found that people's strategies for exploring visualized data change as a function of their spatial abilities. When navigating a 3D interface for semantic document relationships, recall was correlated with spatial ability whereas precision showed a negative correlation. 
Vicente et al.~\cite{VicenteEtAl1987} found that people with low spatial ability took much longer to navigate hierarchical file interfaces and had different navigation patterns compared to subjects with high spatial ability. 
These performance differences persisted (at a reduced magnitude) for interfaces specifically designed to account for an individual's ability level~\cite{VicenteWilliges1988}.
Many studies of spatial ability emphasize people's abilities to make sense of more complex design choices, such as using 3D~\cite{VelezEtAlVis2005} or animation~\cite{CohenHegarty2007}, finding that people with high spatial ability can generally make better use of such techniques.
These differences may extend to different kinds of reasoning judgments: participants with higher spatial ability more easily use visualizations for Bayesian reasoning than individuals with lower spatial ability, and exhibit substantially greater correctness~\cite{OttleyEtAlVIS2015}.  

However, spatial ability is correlated with other factors that may affect visualization performance. The education literature demonstrates that there are substantial differences in spatial ability and other cognitive abilities between different professions (see \citenum{WaiEtAl2009} for a review). 
Moreover, VanderPlas \& Hofmann~\cite{VanderPlasHofmann2016} found that performance on line-up tests amongst undergraduate students is correlated with both visual ability and 
whether or not a student is a STEM major. Froese et al.~\cite{froese2013evaluation} found that training can reduce differences between high and low ability people. Specific fields of training, duration of experience, and other factors associated with a profession, both at the undergraduate level~\cite{yoon2017exploring} and beyond~\cite{WaiEtAl2009}, may all similarly influence people's abilities to use visualizations. Given that different professions exhibit differing average spatial ability~\cite{WaiEtAl2009} and that performance on visualization tasks generally correlates with spatial ability, different professions may exhibit differing performance on basic visualization tasks. We test this conjecture and provide an initial exploration of cognitive differences between professions as they relate to visualization.

\subsection{Visualization Design}
The literature on visualization design (e.g., \cite{Munzner2009,SedlmairEtAlTVCG2012,KirbyMeyer2013,EdgeEtAlTVCG2018,SimonEtAlEuro2015,HallEtAlTVCG2020,McKennaEtAl2014}) emphasizes exploring different domains to inform design processes, asserting that domains might differ in their data, processes and challenges, but not in terms of the cognitive traits of their constituent members. For example, Munzner's Nested Model~\cite{Munzner2009} identifies key threats to the visualization design process and strategies to mitigate these threats.
This model provides guidance for comprehending domain needs and evaluating how well a visualization addresses them; however, understanding the unique cognitive features of a target user group is not cast as a starting point for visualization design. Kirby and Meyer's characterization of visualization collaborations~\cite{KirbyMeyer2013} explicates the critical role of cognitive and perceptual psychologists in system evaluation, but overlooks the potential for domain-specific cognitive and perceptual studies to formatively guide collaborative design. The reflective orientation of these models means that discovery of such differences might not occur until evaluating a fully-implemented design, potentially leading to missed opportunities and ineffective design choices.

To combat such hindsight discoveries, design frameworks typically require developers to explicitly characterize the needs and requirements of users early in the design process. 
Design Study Methodology \cite{SedlmairEtAlTVCG2012} advocates that people understand the visualization literature pertaining to a problem before engaging in problem-driven visualization work, presumably including cognitive and perceptual work related to target data or tasks. However, this connection is limited to generalized perceptual knowledge and is distinct from discovering ``the practices, needs, problems and requirements of the domain experts'' \cite[p. 2436]{SedlmairEtAlTVCG2012}. The separation of user ability and domain knowledge
neglects the possibility for the cognitive traits of domain experts to inform visualization design. Expanded methods~\cite{SimonEtAlEuro2015} recommend employing liaisons (essentially domain-vis intermediaries) to translate domain processes and knowledge between visualization practitioners and domain users.
While including liaisons may implicitly integrate some aspect of cognitive traits (e.g., differing capacities of individuals to use certain representations), liaisons do not account for the possibility of systematic cognitive differences between developer and domain.

The Activity Typology for Visual Analytics recommends developers characterize domain experts to enable ``\textit{Portable Analysis.} The ability to transfer analytic work across people, places, time, and devices'' (emphasis in original text) \cite[p. 271]{EdgeEtAlTVCG2018}. While cognitive variations between disciplines could limit such transferability, understanding the relationship between these variations and related visualization task performance between disciplines could also enrich design by helping characterize the disciplinary capabilities of target groups. 

Recent frameworks provide more direct opportunities to investigate cognitive differences in formulating a design, although these opportunities are often implicit. 
For example, the \textit{understand} activity within the Design Activity Framework \cite{McKennaEtAl2014} requires developers to understand a target domain and its users. While it emphasizes the role of tasks and workflows, the understand activity could 
include cognitive trait assessments across a discipline. Design by Immersion \cite{HallEtAlTVCG2020} offers a collaborative, transdisciplinary approach to problem-driven visualization in which a visualization practitioner's personal engagement, participation, and experience with a target domain informs the design. While Design by Immersion reflects on differences in knowledge, language and tasks, it does not capture potential cognitive differences between visualization practitioners and members of a target domain, instead offering the opportunity to implicitly assess such relationships through longitudinal collaboration with individual experts.

Explicitly accounting for cognitive traits in visualization design methodologies adds more layers of complexity to an already intricate process. While we have preliminary evidence for differences in cognitive abilities across disciplines \cite{yoon2017exploring,WaiEtAl2009}, for differences in visualization use across abilities \cite{VelezEtAlVis2005}, and for the role knowledge plays in shaping visualization use and interpretation \cite{xiong2019curse}, alternative work~\cite{padilla2018decision} suggests that data interpretation across disciplines may be more consistent than indicated by prior studies. We conducted a formal study across disciplines that conventionally reflect a range of spatial ability levels~\cite{WaiEtAl2009}---Education, Computer Science, and Chemistry---to investigate the interplay between disciplinary and cognitive factors in visualization interpretation in order to explore the need for understanding domain-specific cognitive traits in visualization design and evaluation.

\section{Hypotheses}
\label{sec:hypothesis}

Inspired by the knowledge gaps identified in the previous work, and the potential importance of domain-level differences when designing and interpreting empirical evaluations of visualizations, we 
carried out a study of spatial abilities across domains of expertise. 

For tractability, we focus our investigation on computer scientists, chemists, and educators. Computer scientists were an important population to include as they represent “standard  participants” for many studies on task performances and design choices (e.g., ref.~\cite{ChenCZerwinski1997,Chen2000}). In turn, performance differences between Computer Science and other disciplines could have broad implications (e.g., in terms of generalizability of previous work). We chose chemistry because: 1) there is the rich literature exploring relationships between spatial ability and chemistry performance (see ref.~\cite{HarleMarcy2011} and references therein), and 2) the research team had substantial experience working with chemists. Education was selected given the team's experience engaging with educators, and the critical role educators play in developing skills relevant to visualization (e.g., data literacy).  Importantly, these disciplines, on average, reflect different levels of spatial ability. The physical sciences and computer science exhibit higher than average abilities, and educational professionals lower \cite{WaiEtAl2009}. However, these broad categorizations neglect how both training and experience may shape visualization use and how individual abilities may vary within each discipline. Our study investigates three primary hypotheses about the relationship between disciplines, spatial abilities, and visualization use: 

\vspace{3pt}
\noindent \textbf{H1: Spatial abilities will differ across disciplines.}
\noindent Prior studies~\cite{WaiEtAl2009,SheaEtAl2001} in educational psychology show that people's spatial abilities vary systematically across disciplines. Drawing on  previous findings, we anticipate that chemists will have the highest average spatial abilities and educators the lowest. Measuring whether the relationship between spatial ability and professionalization results from systematic preferences (i.e., people of higher spatial abilities are drawn to chemistry) or are learned through training and experience is beyond the scope of this study. Nevertheless, a correlation between ability and discipline indicates fundamental differences in cognitive traits across disciplines that may impact visual reasoning processes. 

\vspace{3pt}
\noindent \textbf{H2: People's abilities to compare visualizations will correlate with their spatial abilities.}
\noindent People's search strategies and abilities to accomplish
visual reasoning tasks, including how well they use particular kinds of visualizations, vary with spatial ability \cite{VanderPlasHofmann2016,OttleyEtAlVIS2015,VicenteEtAl1987,ChenCZerwinski1997,CohenHegarty2007,VelezEtAlVis2005}. People with high spatial abilities tend to use visualizations more effectively. We anticipate these patterns will replicate on an expanded set of visualization tasks, further emphasizing the importance of individual differences in design. 

\vspace{3pt}
\noindent \textbf{H3: People's abilities to compare visualizations will correlate with their disciplines such that average task performance will vary with discipline.} 
\noindent Problem-oriented visualization design methods like design studies \cite{SedlmairEtAlTVCG2012} or design-by-immersion \cite{HallEtAlTVCG2020} assume that disciplinary needs are core to the ``right'' solution to a visualization problem. We anticipate that these needs extend deeper than the data to encompass the abilities of the user directly. Discipline-specific training and experience can also build expertise and familiarity with particular visualization designs and tasks. For example, chemists have spent centuries experimenting with different strategies to represent molecular structures\cite{CookeOrgBioChem2004}, and have explored techniques to represent multidimensional information on a 2D page, both molecular structures\cite{NewmanJChemEd1955} and multidimensional functions\cite{LevinePhysChem}. Instruction on how to use chemical visualization permeates chemical education and corresponding texts\cite{Louden}. When giving advice to potential authors, the editors at a prominent physical chemistry journal noted ``well-composed and scientifically accurate figures constitute the core of a scientific paper" \cite{Kamat2014}. For disciplines like chemistry with a strong emphasis on visuals and systematized visualization education, we anticipate the interrelation of these factors and disciplinary differences in spatial abilities could lead to even higher performance. For example, it is reasonable to expect that chemists will perform better on tasks involving 2D projections of multi-dimensional datasets. 

Confirming these hypotheses would indicate that effective visualizations need to consider not only the needs of the discipline but the abilities of the individuals within that discipline. Such results would reveal significant limitations in the idea of the ``universal'' user.

\section{Study Design}
\label{sec:study}

To evaluate our hypotheses (Section~\ref{sec:hypothesis}), we measured: 1) spatial abilities using the Revised Purdue Spatial Visualization Test: Visualization of Rotations (PSVT:R), a psychometric instrument requiring participants to match different views on the same object through mental rotations~\cite{yoon2011,yoon2011Thesis}, and 2) people's ability to assess correspondences 
between different visualizations.
Inspired by Just Noticeable Difference tasks from vision research \cite{ChangEtAl2016}, we selected the following visualization tasks where participants assess whether visualizations correspond to the same dataset at varying levels of dataset difference (difficulty).
\begin{description}[topsep=2pt,parsep=0pt,itemsep=3pt]
    \item[Pie Chart Correspondence] Can participants determine whether two pie charts contain the same data? 
    \item[Isocontour Correspondence] Can participants recognize whether a 2D isocontour projection matches a given 3D surface?
    \item[Scatterplot Correspondence] Can participants assess which face of a 3D scatterplot corresponds a provided 2D projection?
\end{description}
The above visualization tasks share a common theme: can people reconcile different views of the same data? We chose this theme as it aligns with the emphasis on mental rotation in the PSVT:R and provides a complex
task related ecologically valid scenarios (e.g., small multiples). 
Participants use complex visualizations to reason over high-level tasks that are relevant across disciplines. These tasks draw on elements of prior studies of visualization use and spatial abilities, such as search \cite{ChenCZerwinski1997} and classification \cite{VanderPlasHofmann2016}, extending the findings to a more complex suite of designs and broader set of individual characteristics. We chose to consider isocontour plots since chemistry (one of our target disciplines) often leverages these plots to discuss chemical processes \cite{Levine}, and chemistry students are trained to use them. Similarly, the 3D scatterplot task aligns with the common chemistry task of projecting 3D molecular structures to 2D representations \cite{NewmanJChemEd1955}. We explored the above correspondences by creating a study for online deployment, consisting of a series of tasks for the participant to complete in sequence. The structure and setup of the study are detailed below; see Supplementary Material (SM) for snapshots of the study system. Our study code (stimulus generation, system, and analysis) and data are available on OSF\cite{osfRepo}.

\subsection{Recruitment}
We recruited participants from the different professions by circulating scripted recruitment emails on appropriate professional mailing lists and with institutional contacts (e.g., heads of departments), asking that they share the study with their faculty and graduate students. We also posted to relevant professional social media groups (e.g., LinkedIn, Facebook). In both cases, we informed people that they were free to forward the study to their contacts. We also leveraged our personal networks to initiate snowball sampling. To reduce variance in participant training, we restricted our recruitment to the USA and Canada. Participants were recruited to complete the tasks in a single session. 

\subsection{Study Structure}

After participants agreed to the study consent form, the study process consisted of the following components, as depicted in Fig.~\ref{fig:teaser}.
Participants confirmed readiness to begin the next block at each step, to allow for breaks as needed. There was a final screen where participants could provide information to receive compensation (\$20 gift card). The study results and personal information required for compensation were stored in two separate unlinkable databases to preserve participant anonymity.

During piloting, it became clear that participants perceived the PSVT:R as the most difficult portion of the study, so we kept this test at the start of the study to ensure a more consistent experience across participants. Participants were provided with the standard instructions for the PSVT:R. For the Pie Chart, Isocontour, and Scatterplot tasks, participants were provided with instructions on how to complete each task. They were then presented with a training trial for that task, and provided feedback about their correctness on the training trail. 

\begin{figure}[t]
    \centering
    \includegraphics[width=0.85\columnwidth]{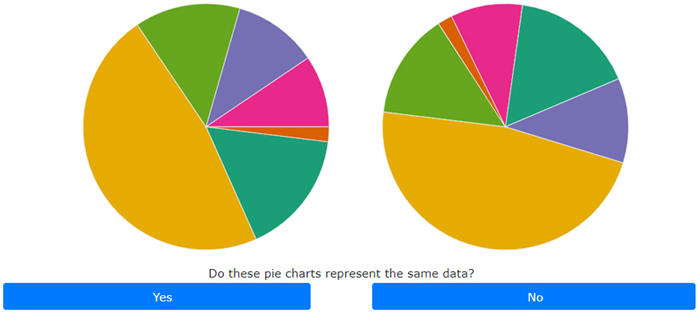}
        \vspace{-1em}
    \caption{Example of the Pie Chart Task. Participants completed 30 trials.}
    \vspace{-3mm}
    \label{fig:pie}
\end{figure}

\subsubsection{Measures} 
For each trial in all blocks, we collected the selected answer and completion time. Additionally, for the interactive tasks (Isocontour and Scatterplot), we also logged all mouse-based interactions in the study window. At the end of the PSVT:R and each task block, participants provided a self rating of their confidence on the previous block using a slider (Low--High) which was converted into a [0--100] scale. 

\subsubsection{Screening and Demographic Survey}
Following the consent form, each participant completed an introductory 
survey to collect information about their profession, education history, visualization usage, participation in pastimes known to relate to spatial ability (e.g., video games), and demographic information. A second stage of screening for color vision acuity was carried out using four Ishihara plates~\cite{Ishihara1972}, as several of our tasks relied on color discernment.

Only people who self-identified as a member of one of our target professions, had finished their undergraduate studies, and correctly answered all Ishihara plates, were permitted to continue the study. A failure to complete the plates correctly may be indicative of color-vision deficiencies or attributable to display conditions, so participants were not informed the reason for being screened out of the study.

We chose to exclude individuals who had yet to complete their undergraduate training as substantial shifts in discipline can take place during that time (e.g., people switching majors). Participants could only choose one discipline when selecting primary field (i.e., CS,  Chemistry, OR Education), and our categorization of participant profession relies on this self-identification; we did not assign people to professional groups based on their education due to mapping challenges. We adopted this approach because our primary focus is assessing differences between disciplines (e.g., Education vs. Chemistry) rather than differences within individual disciplines (e.g., Educators). Exploring within-discipline variations could be especially relevant to Education where individuals often complete undergraduate training in one field and then specialize in education (see SM for a breakdown for our participants). For reference, we focused on recruiting participants with formal education training, and so ``education'' is thus a well-defined professional group for the purposes of this study.

We asked participants to complete the study using a computer, and excluded ones who attempted to complete the study on a mobile device, to increase consistency of display and interaction across participants.

\subsubsection{PSVT:R}
Participants then completed the PSVT:R \cite{yoon2011,yoon2011Thesis}, a thoroughly researched spatial ability test ~\cite{yoon2011Thesis,Maeda2013,MaedaYoon2013,MaedaYoon2016,YoonMetApp2016}. The test is a standardized set of 30 rotation-based spatial ability tests in which participants match analogous rotations of 3D models. More specifically, participants are shown an initial view and and a subsequent rotated view of on asymmetric block. They are then provided with a second block, and must select from a set of distractors what the second block would look like if it were rotated the same way as the first. The test is image-based with no possibility of interaction.

\begin{figure}[t]
    \centering
    \includegraphics[width=0.9\columnwidth]{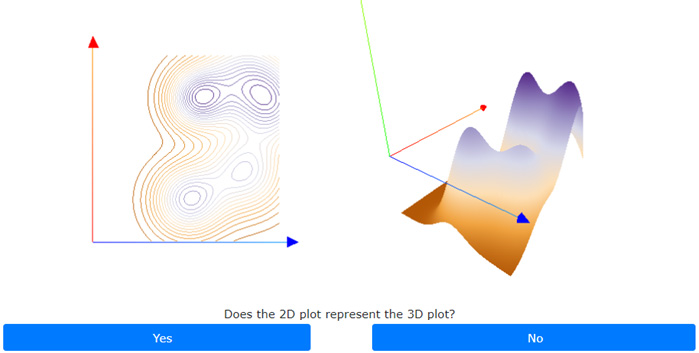}
    \vspace{-1em}
    \caption{Example of the Isocontour Task. The contour on the right starts axis-aligned and can be interactively rotated within a limited range. Participants completed 20 trials.}
    \vspace{-3mm}
    \label{fig:isocontour}
\end{figure}

\subsubsection{Pie Chart Task} Each participant was sequentially shown 30 pairs of pie charts. All participants viewed the same 30 pairs of pie charts in random order. For each pair, participants had to decide whether the pie charts corresponded to the same dataset (\textit{Do these pie charts represent the same data? Yes/No}, as seen in Fig.~\ref{fig:pie}).

We created each pair of pie charts by randomly generating a pie chart consisting of 6 slices (random slice size of 10 degrees or greater, and random order), and then performing various perturbations to that pie chart (or lack thereof) to obtain the second pie chart for the stimulus pair while maintaining the first chart's color labelling. We used a consistent, approximately isoluminant color palette, avoiding visual pop-out effects between specific pairs of slices. The stimulus set contained five pie chart pairs for each of the below perturbation scenarios. For this discussion, \textit{order} refers to the circular order of slices in a pie chart, and \textit{orientation} is the angle a slice makes with the vertical direction. Rotating a pie chart changes its orientation, but not its order. We layered pie chart rotation and changes in slice order on top of data changes to capture scenarios that might more heavily rely on spatial ability; assessing data similarity in the presence of these additional non-data variations involves mental rotations to a potentially great extent. Combining these factors together we arrived at six variants for the Pie Chart Task:

\begin{description}[topsep=2pt,parsep=0pt,itemsep=3pt]
    \item [Identical] Charts were completely identical, having the same slice sizes, orderings, and orientations.
    \item [Identical (Order Conserved, Rotated)] Charts were identical, having the same slices sizes and ordering, but the second pie chart was a rotated version of the first chart.
    \item [Identical (Reordered)] Charts were identical, having the same size slices, but the order of the slices was different between the two. 
    \item [Different (Order and Orientation Conserved)] Charts were different. The second pie chart had the same order and orientation of slices as the first pie chart, but two slice sizes had been altered.
    \item [Different (Order Conserved, Rotated)] Charts were different. The second pie chart had the same order slices as the first pie chart but rotated, and two slice sizes had changed in size.
    \item [Different (Order Not Conserved, Rotated)] Charts were different. The second pie chart did not have the same order of slices as the first chart, and two slice sizes were changed.
\end{description}

\begin{figure}[t]
    \centering
    \includegraphics[width=0.85\columnwidth]{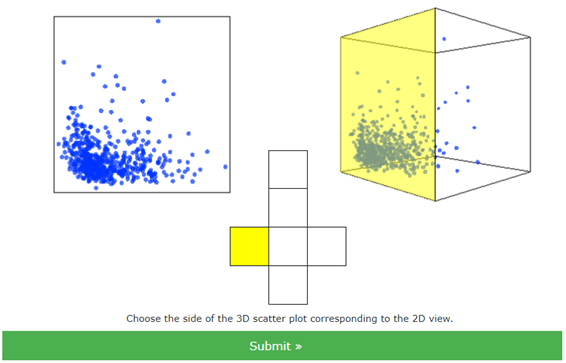}
    \vspace{-3mm}
    \caption{Example of the Scatterplot Task. The 3D cube starts aligned to a random incorrect face and can be interactively rotated within a limited range. Linked highlighting supports face selection. Participants completed 10 trials.}
    \vspace{-3mm}
    \label{fig:scatterplot}
\end{figure}

\subsubsection{Isocontour Task} 
Each participant was sequentially shown 20 pairs of plots. Each pair  consisted of one isocontour plot and one 3D surface plot, both of which used the same color ramp. All participants viewed the same isocontour-surface pairs, but in different random orders. For each pair, participants had to decide if the two charts corresponded to the same dataset (\textit{Does the 2D plot represent the 3D plot? Yes/No}, as shown in Fig.~\ref{fig:isocontour}).

For each 3D surface plot, we gave participants a randomly-generated initial viewing angle of the surface in profile at a slight acute angle with respect to the x-y plane. More specifically in spherical polar coordinates, the initial viewing angle consisted of: 1) a randomly-generated rotation around the vertical direction (i.e., the $z$ axis) such that $\phi \in [0^{\circ},360^{\circ})$, and 2) a second randomly-generated rotation from the $z$ axis such that $\theta \in (70^{\circ},110^{\circ})$. The rotations were performed with respect to the x-y centroid of each surface. Note that each participant saw each 3D surface plot from the same initial view. During each test (i.e., stimulus pair), participants were able to freely rotate the 3D surface about the vertical direction, and could rotate the viewing angle off of the x-y plane by $\pm30^{\circ}$, that is to say $\theta \in (60^{\circ},120^{\circ})$. As a result, a participant could not precisely align the surface plot with the 2D isocontour, and mental rotation was in principle required to relate them.

We created each pair of plots by randomly generating a surface $z=f(x,y)$ composed of five Gaussian peaks on a flat plane (random both in peak position and size) for the isocontour plot, and then performed various perturbations to the initial surface (or not) to arrive at a second surface for the 3D surface plot. At the end of the generation procedure, all surfaces were normalized so that their vertical range was the same. The stimulus set contained five of each of the below types of isocontour-surface plot pairs. In addition to perturbing (or not) the positions of the peaks to arrive at the 3D surface, we also created some stimuli where the 3D surface was inverted (x,y,z) $\rightarrow$ (-x,-y,z) with respect to the isocontour surface, adding additional complexity and the potential for a false friend (a pair of plots that look potentially similar if one does not properly attend to the axes and their relation to gradations).

\begin{description}[topsep=2pt,parsep=0pt,itemsep=3pt]
    \item [Identical]   The data sets were identical. 
    \item [Different (Shifted Peaks)] Two of the peaks in the underlying surface for the isocontour dataset were displaced to arrive at the dataset for the 3D surface plot. 
    \item [Different (Flipped)] The isocontour dataset was flipped along the y=x plane [i.e., (x,y,z) $\rightarrow$ (-x,-y,z)] to arrive at the dataset for the 3D surface, yielding two different surfaces with a symmetric relation. 
    \item [Different (Flipped, Shifted)] The isocontour dataset was flipped along the y=x plane [i.e., (x,y,z) $\rightarrow$ (-x,-y,z)] and then two of the peaks were shifted to arrive at the dataset for the 3D surface.  
\end{description}
 
\subsubsection{Scatterplot Task}
Each participant was sequentially shown 10 pairs of 2D and 3D scatterplots. In each case, the 2D scatter plot corresponded to an orthographic view along one of the faces of the cube corresponding to the 3D scatterplot. For reference, the 3D datasets used for this task were generated by selecting sets of three dimensions from existing multi-dimensional datasets~\cite{ml-data}, and visually inspecting
the datasets to ensure that the selection resulted in an asymmetric data distribution such that orthogonal views were visually distinct.  The participant was tasked with selecting the 3D scatterplot face that corresponded to the 2D view (see Fig.~\ref{fig:scatterplot}). Importantly, a participant's initial viewing angle always aligned with the normal of a cube face other than that corresponding to the 2D scatterplot view, and the participant could only rotate around the face normal by 45\textdegree. As a result, a participant could not exhaustively search for the 2D view by rotating the cube, and mental rotation was in principle required to relate the 3D and 2D scatterplots. Note that the asymmetric nature of the datasets meant that the front and back sides of the 3D scatterplots were distinct, non-superimposable views. 

The front face for a particular 3D scatterplot was the same for each participant, and the exclusion of the front face meant that the 2D view could only be one of the five remaining faces (though participants could select all six faces). In turn, the 2D view for each dataset was systematically varied across participants by professional group such that each of the five faces was equally balanced across that profession. The five possible face selections were also balanced within each participant with two datasets having 2D views corresponding to each possible face selection: top, bottom, left, right, and back (2 $\times$ 5 = 10 pairs of 2D and 3D scatterplots). The dataset order was randomized per participant.

\subsection{Strategies Questions}
Previous work~\cite{STIEFF2007} has analyzed people’s self-reported perspectives on their own practices to probe for discipline-specific approaches to tasks. Similarly, we asked participants to reflect on the strategies that they used when completing the PSVT:R test and each of the task blocks using free-form text. Thumbnail images were shown to remind participants of each task. We put this as the end of the study to avoid priming affects as participants went through the study. Note that the quality of the self-reported strategies likely depends on a participant's capacity for meta-cognition, and the extracted descriptions could be enhanced through observation and elicitation techniques. Such additional investigations were not amenable to our large online study.

\subsection{System Details}
We implemented a browser-based system in JavaScript. Pie charts, isocontour plots, and 3D surface plots were rendered with D3 \cite{D3}, Three.js \cite{ThreeJS}, and delaunator \cite{Delaunator}. The Camera-Controls library ~\cite{Camera-Controls} was used to provide click-and-drag functionality for rotating the 3D plots. Our system code is available on GitHub~\cite{GitRepo} and via OSF~\cite{osfRepo}. 

\subsection{Participants}

Due to the high educational expertise required for valid participants, we were only able to recruit 64 valid participants within the time-frame of the study. The number of participants varied per discipline: 33 Chemistry, 19 Computer Science (CS) and 27 Education.
We note that even though our samples are not uniform, our results still provide valuable insights: there is no magic number for participants required in a study~\cite{Bacchetti2010} and visualization studies often have small numbers of participants with relevant results~\cite{besancon:hal-03012861,Caine:2016:LSS}. And when it comes to statistical evidence, confidence intervals (CIs) with just a hand-full of participants can still provide evidence of differences~\cite{Dragicevic:2016:FSC}. 

The gender statistics across the three disciplines were as follows. Chemistry: 58\% female and 42\% male. CS: 16\% female, 74\% male, 5\% self-disclosed, and 5\% not disclosed. Education: 81\% female, 15\% male, and 4\% not disclosed. These are consistent with known professional gender differences in North America\cite{StatCanCS,OECDTeach}. The average age ($\pm$ standard deviation) for participants was 29$\pm$7 for Chemistry, 27$\pm$6 for Computer Science, and 37$\pm$9 for Education. We found no evidence of a difference in age for CS and Chemistry, but Education participants tended to be older; the details of these analyses are in SM, demographics supplement. Despite possible age differences, we note that cognitive traits are expected to be stable for adults~\cite{PeckEtAlBELIV2012,LiuEtAlCGF2020}.

\section{Results}
\label{sec:results}

\begin{figure}
    \centering
 \includegraphics[width=\linewidth]{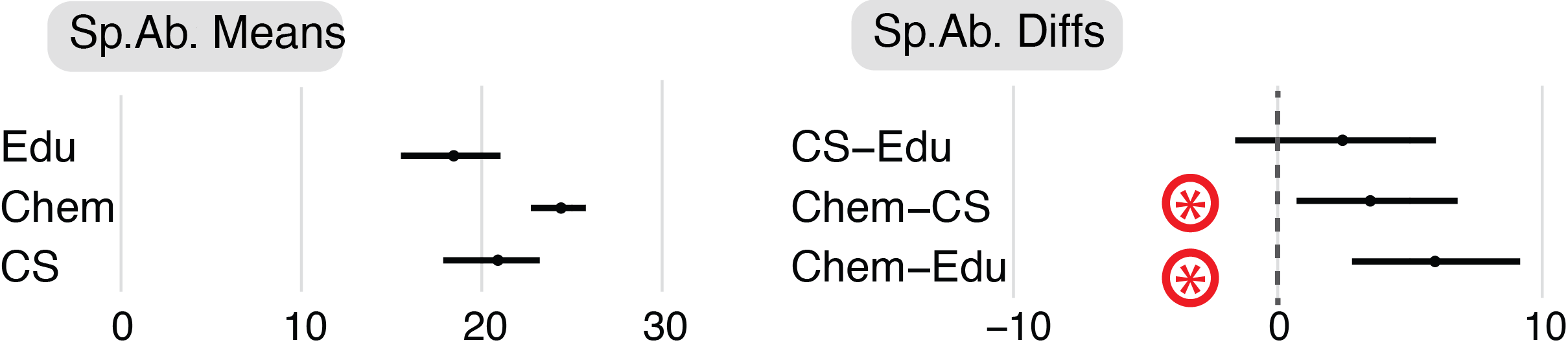}
\vspace{-2em}
    \caption{\emph{H1:} \textbf{Spatial Abilities} per Discipline, with the CI of means (left) and of mean differences (right). In the mean differences plot (right) CIs that are tighter and further away from 0 provide stronger evidence of differences. Red stars indicate evidence of 
     differences between disciplines. Error bars represent 95\% Bootstrap confidence intervals (CIs). }
     \vspace{-3mm}
    \label{fig:H1}
\end{figure}

We analyzed differences in performances and spatial abilities 
using sample means and 95\% CIs,
constructed using BCa bootstrapping (5000  iterations). 
We analyze the CIs using estimation techniques, interpreting them as providing different strength of evidence about the population mean, as recommended in recent reviews \cite{Besancon:2017:SD,Cockburn:2020:TRC,Cumming:2013:NS,Dragicevic:2016:FSC}. 
CIs of mean differences that do not overlap 0 indicate a difference, corresponding to statistically significant results in traditional inferential statistics. 
However, CIs allow for more subtle interpretations: the farther from 0 and the tighter the CI is, the stronger the evidence. Equivalent p-values can be obtained from our CI results following Krzywinski and Altman~\cite{Krzywinski:2013:PoS}. All CIs are calculated using the R \texttt{boot} package \cite{boot-1,boot-2}. Here we report high level findings and mean/mean differences (detailed CIs in SM).  

We analyzed differences in performance across spatial ability using a Pearson's correlation test. When $p < 0.05$ there is evidence of a correlation. We interpret correlation strength using the recommendations by Cohen \cite{Cohen}, interpreting correlations as weak ($R$ between 0.1--0.3), moderate ($R$ between 0.3--0.5), and high ($R$ between 0.5--1.0).

\begin{figure*}[t]
    \centering
    \includegraphics[width = \linewidth]{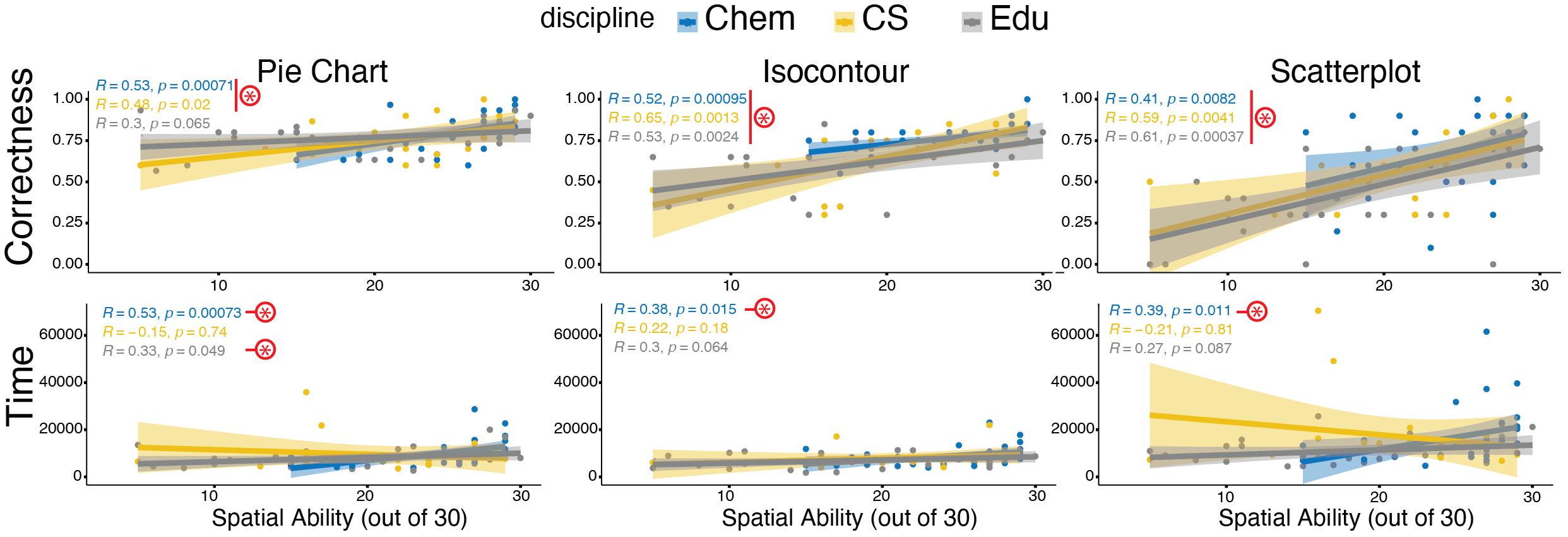}
     \vspace{-2em}
    \caption{\textit{H2:} Spatial Abilities with respect to Performance: Pearson correlation between Spatial abilities and Correctness (top) and Time (bottom), divided by task and colored per Discipline. Red circles indicate evidence of correlation. Time in milliseconds.}
         \vspace{-1em}
    \label{fig:H2}
\end{figure*}

As our study was conducted online without supervision, we 
excluded as outliers participants whose accuracy in each task was below chance (below 50\% for the Pie Chart and Isocontour, and below 16\% for Scatterplots). We interpreted such behavior as a participant not understanding the tasks. We excluded five participants accordingly.

\subsection{H1: Spatial Ability Per Discipline}

To test if participants from different disciplines exhibit different levels of spatial abilities, we followed a procedure similar to Yoon et al. \cite{yoon2017exploring}. 
We calculated each participant's spatial ability as the number of correct trials in the PSVT:R rotation task (score out of 30). We then compared these spatial ability scores across our three discipline groups. 
Our results show (\autoref{fig:H1}) that mean spatial ability was higher for Chemistry (24.4/30), followed by CS (20.9/30) and lowest for Education (18.5/30).

Looking at pair-wise differences of spatial ability, there are differences between Chemistry and the other two disciplines. There is strong evidence that spatial ability for Chemistry was higher than Education (by on average 5.9/30). Chemistry participants also had higher spatial ability than CS participants (by 3.5/30 on average). There may be a slight trend for CS participants to have on average higher spatial ability than Education participants, but  evidence is not conclusive.

While not part of our main hypotheses, we also recorded participants' confidence in the PSVT:R rotation task. 
Chemistry and CS participants tended to be more confident than Education, but our evidence is not conclusive (see SM). 

 \paragraph{$\Rightarrow$} We partially confirmed \textbf{H1}, the spatial abilities of participants differed for some disciplines. The ranking of discipline in terms of spatial ability:  Edu~$\approx$~CS~$<$ Chemistry.

\subsection{H2 - Spatial Ability and Performance}

To determine if spatial ability relates to performance, we looked at correlation between spatial ability and both time and correctness.
We conducted our correlation analysis per task, as their complexity differs.

\textbf{Correctness:}
When looking at each task separately (\autoref{fig:H2}-Top), we see that for all tasks there is evidence ($p < 0.05$) of moderate to high correlation between spatial abilities and correctness for CS and Chemistry
(all $R$ 0.41--0.65). Higher spatial ability leads to higher correctness. For Education, we found
a high correlation  between spatial ability and correctness for the Isocontour and Scatterplot tasks (all $R$ 0.53--0.61), and a trend for the Pie Chart.

\textbf{Time:}
When looking at each task separately 
(\autoref{fig:H2}-bottom), for Chemistry participants, there is evidence ($p < 0.05$) of moderate to high positive correlation between spatial abilities and time to complete the task for all tasks (all $R$ 0.38--0.53). In other words, chemists with higher spatial abilities took longer than those with lower spatial abilities.
We only saw a correlation between ability and time in Education for Pie Charts (again positive), and found no significant correlations with any tasks for CS.

\paragraph{$\Rightarrow$} When looking at performance overall, we partially confirmed \textbf{H2}: participants' correctness 
correlates to spatial ability  (the higher the ability, the higher the correctness) for all tasks for Chemistry and CS participants, and for two of the three tasks for Education (Isocontour and Scatterplots, and there is even a trend for Pie charts). 
However, the correlation between spatial ability and completion times is less clear: for Chemistry participants, higher spatial ability correlates to higher time for all tasks, contrary to our hypothesis.
We did not find a strong correlation between time and spatial ability for CS and Education.

\subsection{H3: Performance Per Discipline}

We hypothesized that performance (time, correctness) will differ depending on discipline, with Chemistry performing best and Education worst (following their spatial ability). We report next our analysis of time and correctness per task (seen in figures) and participants'  self reported  confidence (detailed CIs in SM).

\begin{figure}[t]
    \centering
    \includegraphics[width =\linewidth]{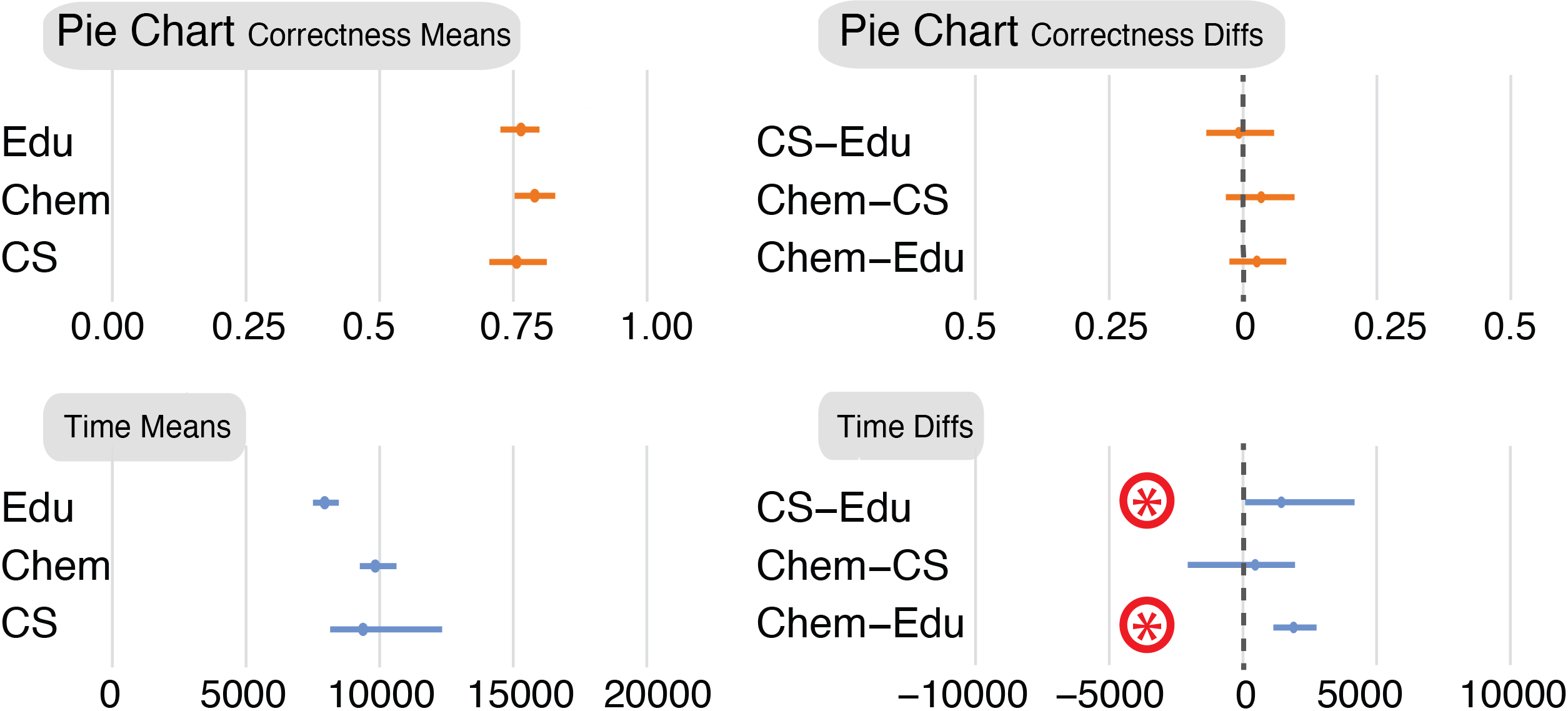}
     \vspace{-2em}
    \caption{\textit{H3-Pie Chart} performance means (on the left) and mean differences (on the right), for Correctness and Time. Time in milliseconds. 
    }
    \label{fig:H3-pie}
\end{figure}

\subsubsection{Pie Charts (\autoref{fig:H3-pie})}

$~~~~$\textbf{Correctness:} 
Mean correctness in the Pie Chart task was higher for Chemistry participants (79.0\%) followed by Education (76.4\%) and then CS (75.6\%). Looking at the mean differences, we found 
no evidence of a difference in correctness between disciplines.

\textbf{Time:} 
Mean times in the Pie Chart task was higher for Chemistry participants (9.8 s), followed by CS (9.4 s) and Education (7.9 s), with
Looking at the mean differences, 
Chemistry was slower than Education (by 1.9 s) and 
CS slower than Education (by 1.4 s).

\textbf{Confidence:} 
Mean confidence was similar across participants for the Pie Chart task (63\% for Chemistry and CS; 61\% for Education).

\begin{figure}[t]
    \centering
    \includegraphics[width = \linewidth]{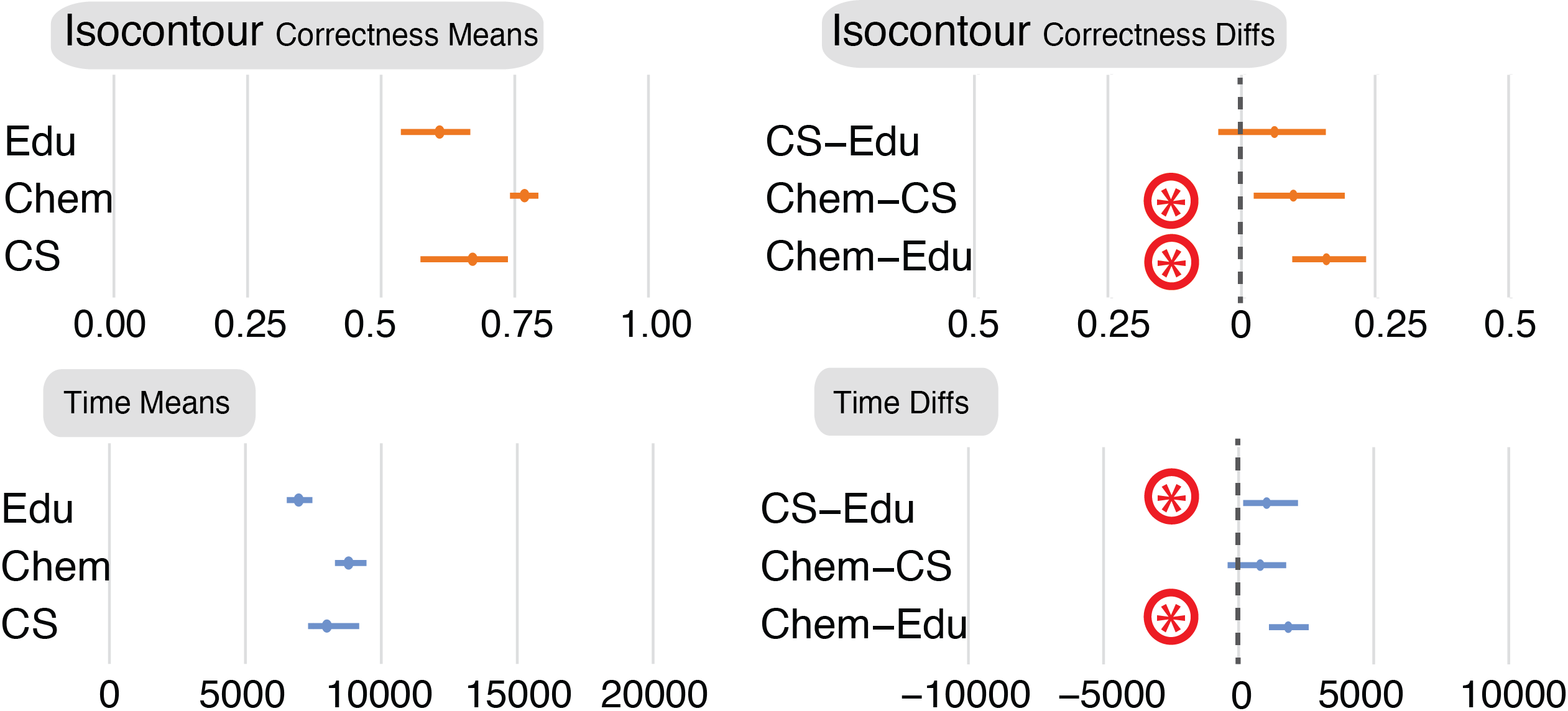}
     \vspace{-2em}
    \caption{\textit{H3-Isocontour} performance means (on the left) and mean differences (on the right), for Correctness and Time. Time in milliseconds.
    }
    \vspace{-3mm}
    \label{fig:H3-iso}
\end{figure}

\subsubsection{Isocontour (\autoref{fig:H3-iso})}

$~~~~$\textbf{Correctness:} 
Mean correctness in the Isocontour task was again higher for Chemistry participants (76.8\%), followed 
by CS (67.1\%) and Education (60.9\%). 
Looking at the mean differences, we found strong evidence of a difference in correctness between Chemistry and the other two disciplines---by 15.8\% compared to Education and 9.7\% compared to CS---but not between CS and Education. 

\textbf{Time:} 
Mean times in the Isocontour task were again higher for Chemistry participants (87.9 s), followed by Cs (79.9 s) and Education (69.6 s). 
Looking at the  mean differences,
both Chemistry and CS were significantly slower than Education (by 1.8 s and 1.0 s respectively).

\textbf{Confidence:} 
Chemistry participants reported the highest confidence (74.8\%), followed by CS (57.8\%) and Education (38.2\%). Looking at mean differences, 
Chemistry participants were more confident than both CS and Education (by 17.0\% and 36.5\% respectively), and CS participants were more confident than Education (by 19.5\%).

\begin{figure}[t]
    \centering
    \includegraphics[width = 0.98 \linewidth]{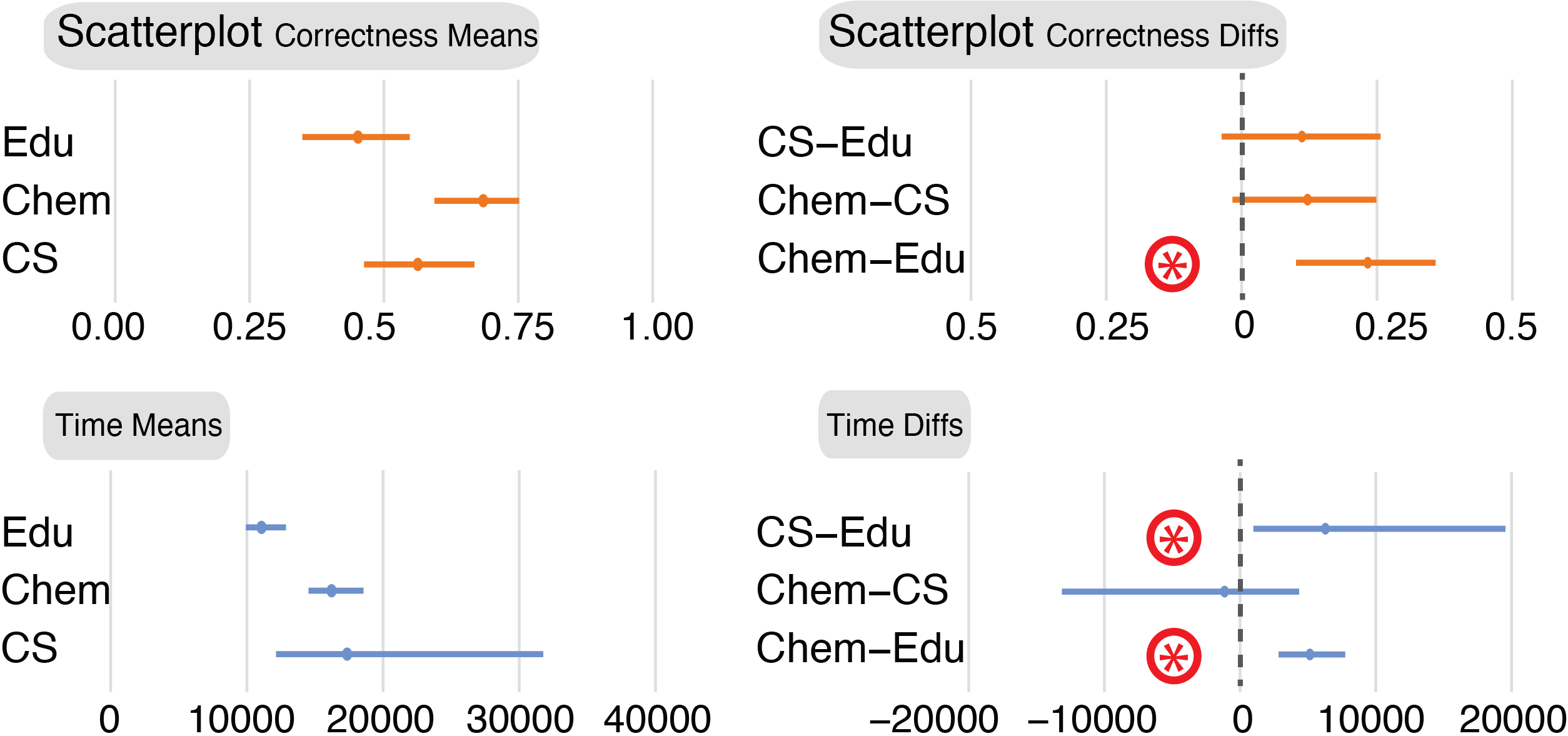}
    \vspace{-1em}
    \caption{\textit{H3-Scatterplot} performance means (on the left) and mean differences (on the right), for Correctness and Time. Time in milliseconds.
    }
    \vspace{-3mm}

    \label{fig:H3-scatter}
\end{figure}

\subsubsection{Scatterplot (\autoref{fig:H3-scatter})}

$~~~~$\textbf{Correctness:} 
The Scatterplot task was the most difficult overall. 
Mean correctness in the Scatterplot task was higher for Chemistry participants (68.4\%), followed by CS (56.3\%) and Education (45.2\%). 
Looking at the mean differences, 
Chemistry was more accurate than Education by 23.3\%. 
Chemistry trended towards being more correct than CS, and CS towards being more correct than Education, but our evidence is not conclusive.

\textbf{Time:} 
Mean times in the Scatterplot task was higher for CS participants (17.3 s), followed by Chemistry (16.2 s) and Education (11.0 s).
Looking at the mean differences, 
both Chemistry and CS were slower than Education (by 5.1 s and 6.2 s respectively).

\textbf{Confidence:} For the Scatterplot task, Chemistry and CS participants reported 
similar confidence (52.9\% and 53.4\% respectively), with Education feeling 
the least confident (36.2\%). Looking at mean differences, both Chemistry and CS participants were more confident than Education participants (by 16.6\% and 17.1\% respectively).

\paragraph{$\Rightarrow$}
Our results partially confirm \textbf{H3}. Chemistry participants tend to perform differently than CS and Education. 
They are more correct than Education participants in two of three tasks (Isocontours and Scatterplots) and were more correct than CS participants for the Isocontour task and likely even for Scatterplots. Chemistry participants also tended to be slower than Education participants across all tasks.

The differences between CS and Education are more subtle. Although mean correctness was higher for CS, 
we found no substantial difference in correctness. But Education participants tended to be faster than CS participants in all tasks. 
Overall, 
we observed a time/error trade-off, with disciplines that are more correct (e.g., Chemistry) tending to also take more time to complete the tasks.

Self-reported confidence generally aligns with performance. Domains do not seem to differ in the Pie Chart, but we see Chemistry  and CS participants being more confident than Education in the Isocontour and Scatterplot tasks. Chemistry participants are also more confident than CS for the Isocontour tasks, mirroring performance. 

\subsection{Strategies \& Qualitative Analysis}
\label{sec:qual}

Table~\ref{tab:survey} provides some highlights from behavioral questions that were asked at the start of the study. The final task of the study was to describe any strategies used for the PSVT:R and each task type. We tabulated all comments, separated by task type and professional group. Two members of the research team did a first pass open-coding on the comments to discover themes and common strategies. They then met to discuss the initial codes and develop a consensus code set.  Finally, one team member conducted a second pass coding on all data using the final code set. Codes were organized into three thematic groups:

\begin{description}[topsep=4pt,parsep=0pt,itemsep=5pt]
\item[Strategies (s)] 13 sub-codes describing different problem-solving approaches (\textit{repeated discrete steps}, \textit{process of elimination}, \textit{use of external aid}, \textit{metaphor or explicit skill transfer}). 
\item[Generated Landmarks (g)] 2 sub-codes describing  mental projection of landmarks: \textit{imagining reference axes} (in the PSVT:R) and \textit{imagining reference angles} (mostly in the pie chart task). 
\item[Visual Landmarks (v)] 9 sub-codes describing  elements mentioned in the strategies (e.g., \textit{angles}, \textit{clusters}, \textit{outliers}, \textit{unique components}).
\end{description}

We rolled up the results across tasks to count participants who were associated with a code at least once. The codes which exhibited the greatest differences are summarized in Table~\ref{tab:strategies}. The complete data and a breakdown by task and profession provided in SM.

\begin{table}[]
    \centering
    \caption{Summary of behavioral questions.}
    \begin{tabular}{p{0.55\columnwidth}p{0.08\columnwidth}p{0.08\columnwidth}p{0.08\columnwidth}} \toprule
         Characteristic & Chem. & CS & Ed. \\ \midrule
         Makes drawings or sketches for work & 79\% & 68\% & 66\% \\ 
         
         \textbf{Strongly Agrees} with ``Drawings and sketches are important to my work.'' & 67\% & 37\% & 33\% \\
         
         Time spent playing video games \newline \phantom{IIII} 0 h/week \newline \phantom{IIII}  $>$5 h/week & 
         \phantom{I} \newline 27\% \newline 24\% & 
         \phantom{I} \newline 16\% \newline 37\% & 
         \phantom{I} \newline 41\% \newline 7\% \\ 
        
        Time spent making visual art \newline \phantom{IIII} 0 h/week \newline \phantom{IIII} $>$5 h/week & 
        \phantom{I} \newline 42\% \newline 6\% & 
        \phantom{I} \newline 37\% \newline 5\% & 
        \phantom{I} \newline  22\% \newline 22\% \\ \bottomrule
         
    \end{tabular}
    \label{tab:survey}
\end{table}

\begin{table}[]
    \centering
    \caption{A selection of codes from task feedback analysis. Percentages represent participants coded at least once, by profession. The prefixes  indicates relations to the three thematic groups in the main text: Strategies (s-), Generated Landmarks (g-), and Visual Landmarks (v-). }
    \begin{tabular}{p{0.37\columnwidth}rrr} \toprule
         Code & Chem. & CS & Ed. \\ \midrule
        s-discrete steps & 33\% & 58\% & 62\%\\
        s-skills transfer & 12\% & 10\% & 0\%\\
        s-used prop or hand & 21\% & 26\% & 50\%\\
        s-used symmetry & 18\% & 0\% & 11\% \\
        s-used mirroring & 21\% & 5\% & 0\% \\
        s-interaction & 76\% & 58\% & 50\% \\
        s-had difficulties & 6\% & 5\% & 27\% \\ 
        g-imagined axes & 12\% & 0\% & 0\% \\
        g-imagined angles & 24\% & 0\% & 8\%\\
        v-tracked unique features & 36\% & 21\% & 15\% \\
        v-angles & 30\% & 26\% & 11\% \\
        v-used visible axes & 70\% & 63\% &38\% \\
 \bottomrule
         
    \end{tabular}
 
    \label{tab:strategies}
\end{table}
\section{Discussion}
\label{sec:discussion}

Our results indicate that both an individual's cognitive abilities and discipline influence how they use visualizations. These differences manifest in both statistical performance and qualitative strategies. 
The relationship between discipline, abilities, and tasks offers several implications for visualization, including new insight into how we might reason about users in problem-driven design. 

\paragraph{Spatial ability, discipline and performance} We explored how both spatial ability and discipline influence task performance (correctness). 
While these two factors are often treated independently, our results reveal a complex interplay. We confirm past findings~\cite{WaiEtAl2009} that spatial ability varies with discipline (Fig.~\ref{fig:H1}), with Chemistry having higher spatial ability than CS and Education 
(\textbf{H1}). We also found moderate linear correlations between task correctness and spatial ability across all three visualization tasks (Fig.~\ref{fig:H2}-top, \textbf{H2}). These relationships were stronger in the Isocontour and Scatterplot tasks. One possible explanation for the greater correlation on these tasks is that they involve mental rotations of 3D visualizations, aligning more strongly with our spatial ability test. Chemistry substantially outperformed both CS and Education on the Isocontour task~(see Fig.~\ref{fig:H3-iso}), and for the Scatterplot task outperforms Education and there is a trend towards outperforming CS. We failed to find notable differences between Education and CS, which aligns with the lack of difference in their spatial abilities (Fig.~\ref{fig:H1}). In contrast, the Pie Chart task involves assessing 2D rotations of planar objects, and is thus more removed from our spatial ability test; here we found no differences in accuracy across disciplines (Fig.~\ref{fig:H3-pie}-top). 

The fact that educators can exhibit differing abilities and performance is key for the visualization community. Teachers are society's knowledge stewards, responsible for guiding the development of regular citizens and future thought leaders alike. Supporting educators should be a top priority for visualization. Visualization literacy is critical for civic participation and leading-edge visualizations may miss wider adoption if 
they are inaccessible or ineffective for a key group that influences society's intellectual spheres from a formative stage.

Beyond correctness, 
we found disciplinary differences in completion times. In general, chemists took longer on tasks than educators (bottom of Figs.~\ref{fig:H3-pie},~\ref{fig:H3-iso}, and ~\ref{fig:H3-scatter}), even on the \pie~ where their correctness was more comparable ~(\autoref{fig:H3-pie}-top). Furthermore, we found moderate positive correlations between completion times and and spatial ability across all three tasks for chemists: chemists with higher spatial ability were slower. We conjecture that this correlation might arise from individuals with higher spatial ability being able to assess the full complexity of the tasks. For example, in their free-form strategy descriptions, some chemists noted the mirror image relationship between opposite sides of the cube during the 3D scatterplot task whereas no computer scientists or educators noted that valid relationship. Critically, we must not confuse discipline capacity or difficulties with time. Designers may need to probe the multifacted processes of a discipline's visualization use, not reducing comparisons to single metrics. 

\paragraph{Strategies and training} Our survey results and participants' strategy comments provide nuanced insights into the performance results. 
For example, reflecting on the strategy codes (Table~\ref{tab:strategies}), all three disciplines described concrete strategies for tackling the Pie Chart task, in contrast to the Isocontour and Scatterplot tasks where the lower performance of educators coincided with vague strategies and comments about having difficulty with the task. Moreover, Pie Charts are relatively common visualizations, and included in many grade school curricula (thus, teachers are specifically trained to teach about them). This familiarity may explain the lack of substantial performance variations 
even though cognitive differences exist between the disciplines. 

However, people reported different strategies across disciplines. For example, some chemists described comparing pie chart slices to characteristic angles like 90\textdegree~in order to assess changes, while only two non-chemistry participant (both Education) reported using generated landmarks to complete any task. Knowledge of such strategy differences (even in the absence of performance differences) could provide a basis for engineering distinct techniques to further enhance each disciplines performance. The absence of performance differences for pie charts coinciding with strategy differences does raise additional questions: Do the strategy differences stem from training? Or do individuals develop different strategies, accounting for their cognitive differences, such that they can comparably engage with common visualizations?  

We found explicit evidence that people's prior training can inform their visualization use via skill transference. Some chemists invoked concepts related to \emph{chirality} (captured in code \textit{s-symmetry} in Table~\ref{tab:strategies}) when describing how they worked through the PSVT:R test and Scatterplot task. Chirality captures the idea that molecules or other objects can be non-superimposable mirror images (e.g., like a person's left and right hands). Chirality and related concepts (e.g., handedness and stereochemisty) span synthetic chemistry \cite{SolomonsFryhle,Louden,CareySundberg} to biochemistry \cite{VoetVoetPratt}, and are an integral part of a chemistry education. The 3D blocks used in the PSVT:R and the scatterplots in our study are asymmetric (chiral) objects and previous education work~\cite{STIEFF2007} has demonstrated that chemists can transfer discipline-specific empirical strategies for addressing molecular problems to non-chemical abstract data and tasks.
This skills transference aligns with chemists' improved performance on the PSVT:R and isocontour tasks. Such skills transference also occurred at a lower level: one computer science participant noted using  ``intuition based on engineering design courses'' when completing the PSVT:R test while other participants compared the isocontour task to navigating a mountain or a topographical map. 

We also see evidence of skills transference in the higher use of props by the Education participants (double the other disciplines). Though gestures can improve spatial reasoning performance, people tend to use such gestures more frequently when they experience difficulties in assessing spatial rotation \cite{chu2011nature}.  Educators are trained to teach students to approach problems from a variety of ways, including concretizing abstract problems~\cite{szendrei1996concrete}. This may point to the potential benefit of data physicalization for some people.  

Skill transference has two possible consequences for design. First, we could look to existing interfaces and visualizations for a target discipline-specific task to guide tools that better encourage skill transference. Alternatively, if we know that a discipline performs better on a task using discipline-specific concepts, then 
that discipline's pedagogy may help visualization designers understand the visuals and metaphors used to introduce the concept and abstract these ideas to support and train other disciplines for whom a given task is difficult.

\paragraph{Disciplines and communication} Participants' open-ended responses also illustrate potential communication barriers during problem-driven visualization design processes 
(e.g., design studies). Van Wijk\cite{VanWijkCGA2006} characterized the challenges during visualization collaborations in terms of knowledge and interest gaps including linguistic barriers (e.g., alternative understandings of the word ``interphase''). Other work \cite{SedlmairEtAlTVCG2012, HallEtAlTVCG2020,McKennaEtAl2014} 
stresses the importance of understanding and using a target discipline's language. While our tasks were discipline agnostic, the linguistic difference we observed across CS, Chemistry, and Education in describing their task completion strategies (see Table~\ref{tab:strategies} and quotes in SM) highlight how individuals from different disciplines may describe a common task/experience differently. For example, on the PSVT:R test, many participants discretized the mental rotation process, but only chemists 
explicitly discussed imagined rotational axes (12\%, \textit{g-imagined axes} in Table~\ref{tab:survey}). This linguistic difference may relate to discipline-specific training: rotational axes are integral to characterizing molecular symmetry, a key concept in chemistry \cite{HousecroftSharpe,Hollas,AtkinsDePaula}. However, our study involves no chemical/molecular information. 
Differences in language around core visualization concepts like tasks, when considered in conjunction with unique needs of a discipline, may pose unexpected barriers to understanding what people need to know about their data (e.g., task abstraction \cite{Munzner2009,SedlmairEtAlTVCG2012,McKennaEtAl2014}). Design processes should pay close attention to how these differences can inhibit effective participatory design.

\paragraph{Demographic influences}
Variations in spatial ability and task performance are qualitatively mirrored by participant behaviours beyond discipline-specific skills (Table~\ref{tab:survey}).  
Chemists are more likely to strongly agree that graphical representations of information are important to their work and to make drawings or sketches for their work. Educators are less likely to play video games compared to CS and Chemistry. If certain behaviors or attitudes coincide with spatial ability, then historical design approaches rooted in understanding a target discipline might indirectly capture some variations in cognitive ability. However, these types of connections are likely nuanced. For example, educators spent more time making visual art than CS or chemists (see Table~\ref{tab:survey}), but exhibit lower spatial ability and task correctness scores. 

\paragraph{Limitations \& outlook}
The goal of our work was to take a first step in shedding light on professional cognitive differences and their implications for visualization. Therefore, we tested basic visualization skills, rather than embedding our study in difficult to control real-world tasks. While people do not regularly perform the exact study tasks, similarity assessments across visualizations are not unusual. For example, our pie chart comparison task maps to a small-multiples scenario where a person is assessing whether they have changed how they spend their days. 

We cannot directly assess the causal origins of discipline variations observed in this study. Previous work~\cite{ChengMix2014,Lowerie2017,SORBY201320,MillerHalpern2013} indicates that spatial training regimes can improve spatial ability and discipline-specific performance/attainment. 
However, the gains from a particular training regime are not necessarily temporally enduring nor generalizable across multiple disciplines~\cite{MillerHalpern2013}. Furthermore, spatial ability interventions appear to have only a small effect on discipline performance~\cite{StieffUttal2015}, and longitudinal analysis \cite{WaiEtAl2009} has demonstrated that individuals with high spatial ability are more likely to pursue careers in particular disciplines. It is an open  question whether professional differences are due to training or the intrinsic capabilities of the people they attract.

It is worth noting that additional factors can influence task performance, such as representational fluency, which has been explored in the context of domain-specific problem solving \cite{MooreEtAl2013,StieffCogInstruct2011} and visualization \cite{Parsons2018}. However, we observed correlations between task performance and spatial abilities for all disciplines, so we anticipate that spatial ability is an important underlying source of variations between the disciplines. 
Natural differences in demographic factors exist across these populations (e.g., our participant groups exhibited differing levels of education attainment, gender balances, and average ages, see demographic SM). While they may contribute to the observed phenomena, several of these variations align with known differences between disciplines (e.g., gender-skewed male in computer science, but female in education). These differences were not controlled in our study and we thus cannot use them to explain our results. Nevertheless they open up intriguing avenues for further studies.

Regardless of the origins of the differences and other potential factors, our work illustrates it is not necessarily valid to treat different disciplines as equivalent in terms of task performance and spatial ability.
Importantly, it is reasonable to conjecture that the observed differences are relatively robust, and that professional differences in task performance and cognitive abilities would endure, with implications for visualization. We did not recruit from individual institutions, but took a broad approach contacting groups of individuals across North America. Our study implicitly reflects a broad set of disciplinary training regimes.

In general, differences in cognitive ability, language, and familiarity with visualizations across disciplines all appear to affect how well people can use visualizations. 
However, the interplay between these factors is complex: neither ability nor discipline alone can explain the quantitative and qualitative differences in our data and further research is warranted.  
By studying visualization use across different disciplines and a range of cognitive abilities, we can begin to catalogue professional differences in cognitive abilities that drive more effective, discipline-specific visualization practices. 
We hope our work will encourage further research into the underexplored space of professional cognitive differences, looking to traits beyond spatial ability.

\section{Conclusion}
\label{sec:conclusion}
In this study, we 
explored visualization task performance as it related to both discipline and spatial ability, providing a new lens for understanding visualization use: professional cognitive difference. Our results provide preliminary evidence that critical differences exist between disciplines, and indicate the need to more deeply consider cognitive, social, and demographic factors in defining effective visualization.
Our work also reveals that disciplines leverage distinct strategies and language to address discipline-agnostic visualization tasks, and these variations may or may not align with differences in visualization task performance and confidence. In general, disciplinary differences involve an intricate combination of social factors, training, and other individual differences that are likely interdependent. To fully empower individuals to understand their data, 
we must assess, respect, and celebrate such differences as we design visualizations. Investigating this rich space will require many additional studies and extensive future work, which we hope our results will inspire.
Our study provides a first foray into this space, opening new avenues to better understanding the ``user'' in problem-driven visualization. Our data and code are available on OSF\cite{osfRepo} to support and inspire future studies.

\acknowledgments{
We acknowledge the support of NSF \#1764089 \& 2046725 and NSERC [RGPIN-2015-03916]. We thank Dr. So Yoon Yoon (University of Cincinnati) for providing the revised PSVT:R materials. 
}

\bibliographystyle{abbrv-doi}

\balance
\bibliography{template}
\end{document}